\documentclass[11pt,twoside]{article}

\usepackage{asp2006}
\usepackage{epsf}
\usepackage{psfig}
\usepackage{lscape}

\begin{document}
\title{The contact system V566 Ophiuchi revisited}   %%% Fill in title
\author{Liakos, A., Varrias, S. and Niarchos, P.}   %%% Fill in author names
\affil{Department of Astrophysics, Astronomy and Mechanics,
National and Kapodistrian University of Athens, GR 157 84
Zografos, Athens, Hellas}    %%% Fill in author affiliations

\begin{abstract} %%% Abstract to run on from here.
New CCD photometric observations of the eclipsing binary V566 Oph
have been obtained. The light curves are analyzed with the
Wilson-Devinney code and new geometric and photometric elements
are derived. A new O-C analysis of the system is presented and
apparent period changes are discussed with respect to possible
Light-Time Effect in the system.
\end{abstract}

\section{Introduction}

The aim of the present study is the investigation for a possible
qualitative connection between the solutions derived from the
light curve and O-C diagram analysis of the system. The presence
of possible tertiary components orbiting the eclipsing binary
(hereafter EB) and the mass transfer process are mechanisms which
can affect both the period of the EB and the shape and brightness
levels of the light curves (hereafter LCs). These two independent
methods of analysis (light curve and O-C diagram analysis) are
applied in the present study and combined with the results from
previous spectroscopic ones in order to get a better and more
clear picture of the system. \\
The system V566 Oph (= HIP 87860 = HD 163611, $\alpha_{2000} =
17^{h}56^{m}52.4^{s},~\delta_{2000} = +04^{\circ}59'15.3''$) is a
W UMa - type eclipsing binary, discovered by \citet{H35}. The
system has been observed photometrically and spectroscopically by
many researchers in the past (for references see \citet{D06}). The
distance of the system has been determined by HIPPARCOS mission to
be 71.531 pc \citep{P97}. \citet{PR06} carried out a period study
of the system and calculated the physical and orbital elements of
a possible third body orbiting the EB. Later, \citet{P06}
published the most recent spectroscopic mass ratio of the system
using the broadening function (BF) method and the rotational
profile fitting. They also determined the spectral class of the
system as F4V, and found no spectroscopic evidence of a tertiary
component.

\section{Observations and analyses}

The system was observed during four nights in May 2009 at the
Athens University Observatory, using a 40-cm Cassegrain telescope
equipped with the CCD camera ST-8XMEI and the B, V, R, I Bessell
photometric filters.

Differential magnitudes were obtained by using the software
Muniwin v.1.1.23 \citep{H98}, while the stars SAO 122955 and GSC
0425-1090 were selected as comparison and check stars,
respectively. Two times of minima have been calculated using the
\citet{KVW56} method and are presented in Table \ref{tab1}.

\begin{table}

\caption{The times of minima derived from our observations}
\label{tab1} \vspace{0.5mm}\centering{ \scalebox{0.83}{

\begin{tabular}{ccc}

\hline \hline
$HJD~(2400000.0+)$         &      $Error$    &    $Type$         \\
\hline
54980.5158                 &      0.0002     &      II           \\
54982.3590                 &      0.0005     &      I            \\

\hline\hline

\end{tabular}}}

\end{table}

\subsection{Light curve analysis}

The light curves (hereafter LCs) were analysed with the PHOEBE
0.29d software \citep{PZ05} which uses the 2003 version of the WD
code \citep{WD71,W79}. Mode 3 was selected for the light curve
analysis and the method of Multiple Subsets was used in order to
obtain the final photometric solution. The value of the
temperature $T_{1}$ of the primary component was adopted according
to the spectral classification of \citet{P06}, and the mass ratio
of the system was initially set at the spectroscopic value
$q_{sp}$=0.263(12) \citep{P06}, but during the final calculations
it was adjusted with fixed step in order to be within the error
limits of the $q_{sp}$. The albedos $A_{1},~A_{2}$ and the gravity
darkening coefficients $g_{1},~g_{2}$ were given their theoretical
values according to the spectral type of each component. The
synthetic and observed LCs and the 3-D model of V566 Oph are shown
in Fig. 1, while the derived parameters from the LC solution are
listed in Table \ref{tab2}.

\begin{table}

\caption{The parameters of V566 Oph derived from the LCs solution}
\label{tab2} \vspace{0.5mm}\centering{ \scalebox{0.8}{
\begin{tabular}{cc| ccccc}

\hline\hline
$Parameter$                 &     $value$    &    $Parameter$    &        \multicolumn{4}{c}{$value$}                    \\
\hline
$q~(m_{2}/m_{1})$           &    0.252~(1)   &                   &      B       &       V      &        R     &    I     \\
\cline{4-7}
$i$~[deg]                   &    80.7~(2)    &   $x_{1}$$^{**}$  &      0.625   &     0.510    &    0.429     &  0.351   \\
$T_1$$^{*}$ [K]             &       6765     &   $x_{2}$$^{**}$  &      0.639   &     0.517    &    0.437     &  0.358   \\
$T_2$ [K]                   &     6650~(5)   &   $L_{1}/L_{T}$   &  0.757~(2)   &   0.766~(3)  &    0.769~(3) &0.770~(3) \\
$A_1$$^*$=$A_2$$^*$         &       0.5      &   $L_{2}/L_{T}$   &  0.207~(1)   &   0.213~(1)  &    0.216~(1) &0.219~(1) \\
$g_1$$^*$=$g_2$$^*$         &       0.32     &   $L_{3}/L_{T}$   &  0.036~(2)   &   0.021~(2)  &    0.015~(2) &0.012~(3) \\
\cline{4-7}
$\Omega_{1}$=$\Omega_{2}$   &   2.295~(3)    &                   &    $Pole$    &    $Side$    &    $Back$    &          \\
\cline{4-7}
$Fillout Fact.$             &     36.7\%     &        $r_1$      &     0.483    &     0.527    &     0.556    &          \\
\cline{1-2}
$\chi^{2}$                  &    0.08404     &        $r_2$      &     0.264    &     0.277    &     0.326    &          \\

\hline\hline

\end{tabular}}}
 \textit{$^*$assumed}, \textit{$^{**}$\citet{VH93}},
\textit{$L_{T} = L_{1}+L_{2}+L_{3}$}
\end{table}

\begin{figure}[h!]
\plottwo{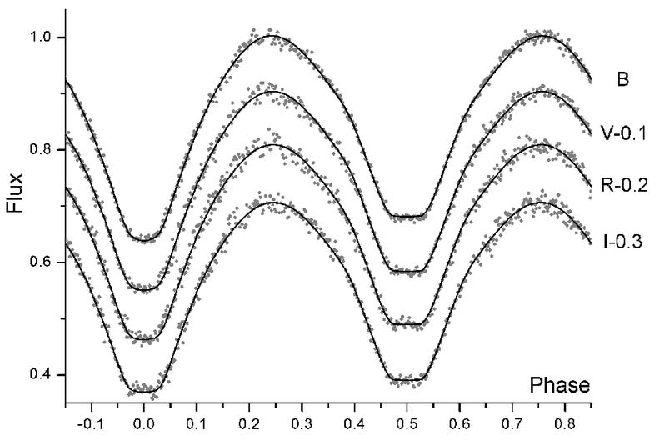}{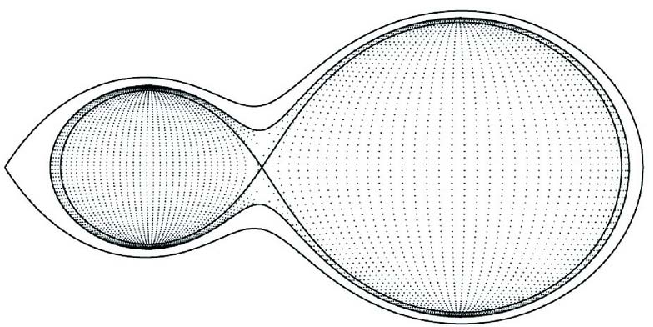}

\caption{Left panel: The synthetic LCs (solid lines) along with
the observed ones (gray points). Right panel: The 3-D model of the
system at the phase 0.75.}

\end{figure}

\subsection{O-C diagram analysis}

The least squares method with statistical weights has been used
for the analysis of the O-C diagram. The weights assigned to
individual observations were set at w=1 for visual observations, 5
for photographic and 10 for CCD and photoelectric observations.
The current O-C diagram of V566 Oph includes 201 times of minima
taken from the literature and two from our observations (see Table
\ref{tab1}). The ephemeris: $Min~I=HJD~2442911.23402 +
0.4096457^{d} \times E$ was used as the initial one for the O-C
analysis of the compiled times of minima. Due to the distribution
of the O-C points, one periodic LITE (\textbf{LI}ght \textbf{T}ime
\textbf{E}ffect) curve \citep{FCH73,I59} and one parabola were
chosen to fit the O-C diagram.  \\
In Table \ref{tab3} the derived parameters are presented, namely:
the mass of each component of the EB, $M_1$ and $M_2$,
respectively, the ephemeris ($Min.~I$ and $P$ for the linear one
and $c_{2}$ for the quadratic one), the period change rate
$\dot{P}$ and the mass transfer rate $\dot{M}$ of the EB, the
period $P_3$, the minimal mass $M_{3,min}$ and the eccentricity
$e_3$ of the third body, the HJD of the periastron passage $T_0$,
the semi-amplitude $A$ of the LITE, the argument of periastron
$\omega$, and the maximum angular separation $\alpha$ between the
third body and the EB. The O-C solution is illustrated in Fig. 2.

\begin{table}

\caption{The results of the O-C diagram analysis for V566 Oph}
\label{tab3} \vspace{0.5mm}\centering{ \scalebox{0.78}{

\begin{tabular}{cccc}

\hline\hline
$Parameters~of~the~EB$                  &     $value$     &     $Parameters~of~the~LITE$      &          $value$         \\
                                        &                 &     $and~of~the~3^{rd}~body$      &                          \\
\hline
$M_1^{*}+M_2~(M_\odot)$                 &  1.4 + 0.35     &          $P_3$~[yrs]              &         62.3~(3)         \\
$Min. I$~(HJD)                          & 2442911.183~(5) &         $T_0$~[HJD]               &     2408118~(1791)       \\
$P$~(days)                              &  0.409646~(1)   &         $\omega$~[deg]            &         2~(14)           \\
$c_{2}~(\times 10^{-10})~[days/cycle]$  &    1.457~(2)    &          $A$~[days]               &         0.013~(3)        \\
$\dot{P}~(\times 10^{-7})~[days/yr]$    &    2.598~(2)    &            $e_{3}$                &         0.6~(2)          \\
$\dot{M}~(\times 10^{-8})~[M_\odot/yr]$ &    9.969~(8)    &      $M_{3,min}~[M_\odot]$        &         0.29~(5)         \\
                                        &                 &         $\alpha$~[mas]            &           279.7          \\
\hline
$\chi^{2}$                              &     0.0521      &                                   &                         \\
\hline\hline
\textit{$^*$assumed}
\end{tabular}}}

\end{table}

\begin{figure}[h!]
\plottwo{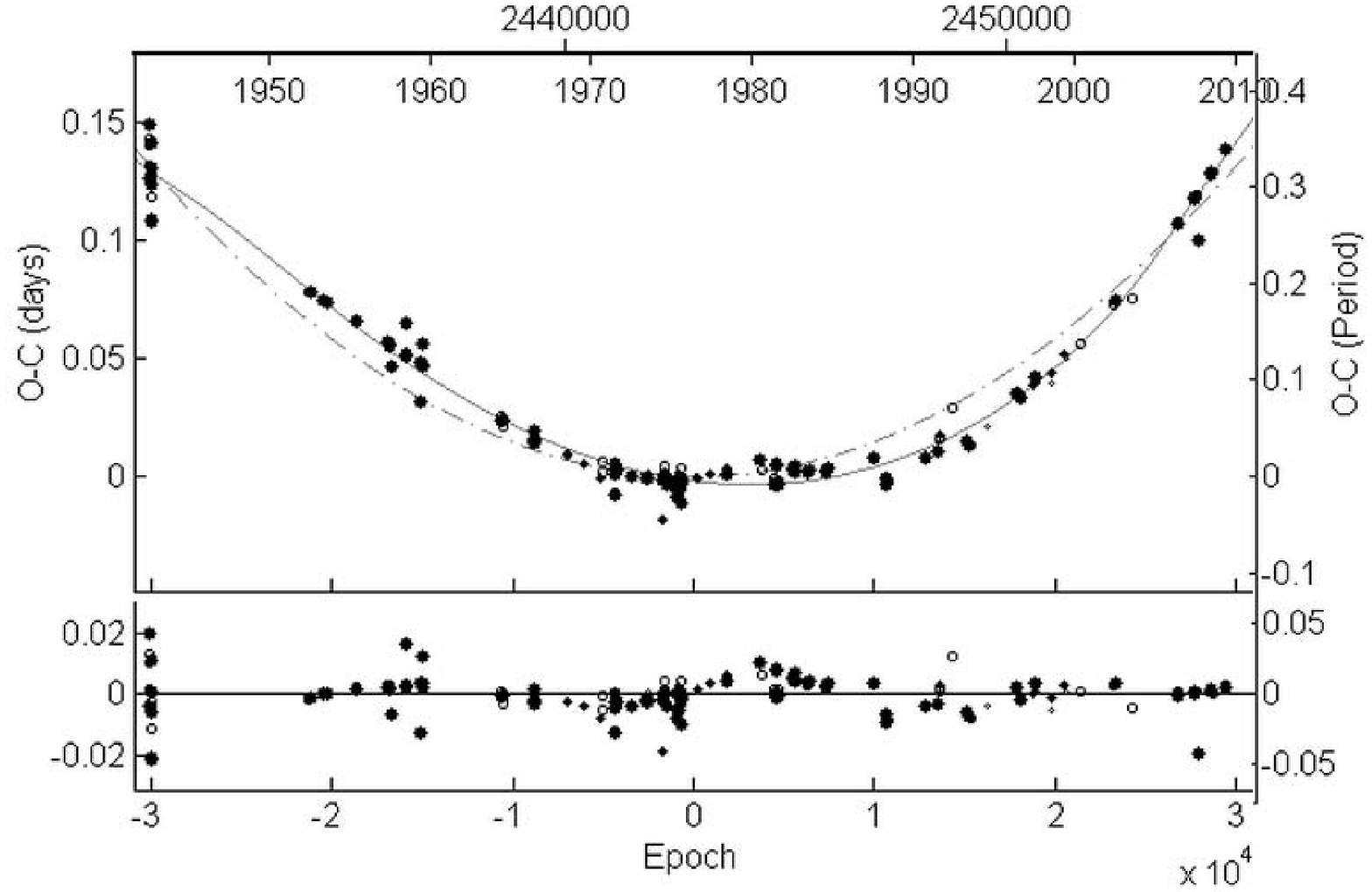}{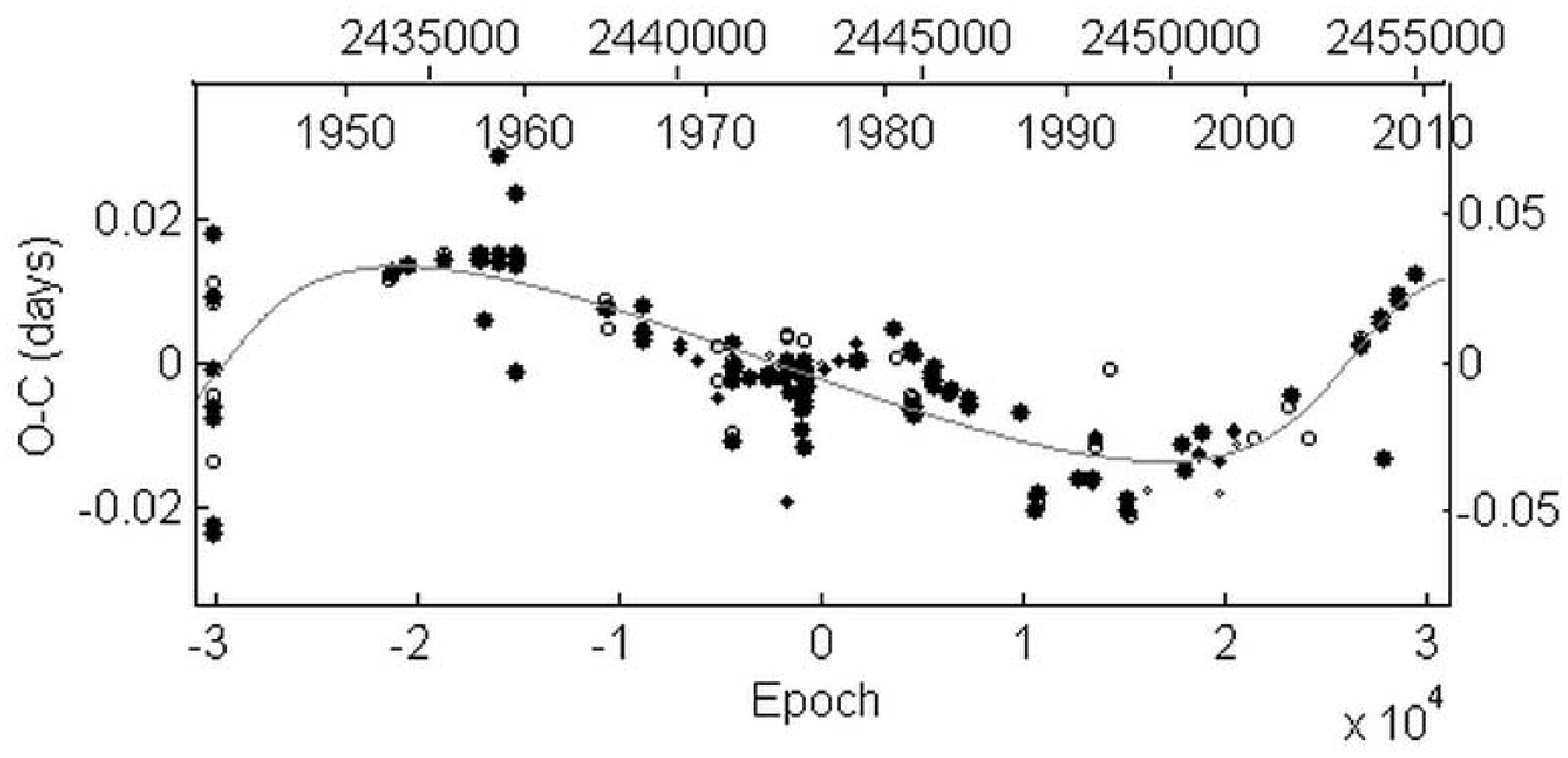}

\caption{Left panel: upper part: The total fit (solid line) on the
O-C points (black points), the parabola (dashed line) and lower
part: the O-C residuals. Right panel: The sinusoidal fit on the
O-C points after the subtraction of the parabola.}
\end{figure}

\section{Discussion and Conclusions}

Complete new BVRI light curves were obtained for V566 Oph. The
results of the LC solution show that V566 Oph is an overcontact
system (fillout factor 36.7\%) not in thermal equilibrium. An
additional light contribution to the luminosity of the EB was
taken into account in the LC solution and it was found 2.1\%. The
periodic variation of the orbital period of the system could be
explained by adopting the existence of an additional component,
which was found to have minimal mass of 0.29 $M_\odot$. Assuming
that all members of the triple system are main sequence stars, we
found that the luminosity from the third body, derived from the
light curve analysis, could be produced by a star with 0.47
$M_\odot$ and an orbital inclination of $i'=41^{\circ}$. Since
such a body was not detected by the spectroscopic observations of
\citet{P06}, the most probable explanation is that it is neither a
main sequence star, nor does contribute to the EB's period
formation. The large angular separation, between the possible
third body and the EB, allows further investigation of its
existence by large interferometers. Moreover, the O-C residuals
show another variation, not strictly periodic, which is possibly
caused by another physical mechanism, such as magnetic activity,
although its signature (O'Connell effect) was not observed in the
LCs. The steady increase rate of its period is probably due to the
mass transfer procedure whose direction of the flow is from the
less massive to the more massive component.

%%% MAIN BODY OF TEXT GOES HERE. CONSULT "INSTRUCTIONS FOR AUTHORS USING
%%% LATEX2E MARKUP", SECTIONS 2.3-2.6 FOR HELP WITH EQUATIONS, FIGURES,
%%% AND TABLES.

%\section{}   %%% Top level section head (remove "%" symbol)
%\subsection{}   %%% Second level section head (remove "%" symbol)
%\subsubsection{}   %%% Lowest level section head (remove "%" symbol)
%\section*{}    %%% Unnumbered top level section head (remove "%" symbol)
%\subsection*{}   %%% Unnumbered second level section head (remove "%" symbol)

%\acknowledgements %%% Text of acknowledgements runs on after this command.

%%% THE BIBLIOGRAPHY
%%%
%%% CONSULT SECTION 3 OF "INSTRUCTIONS FOR AUTHORS" FOR HOW TO USE NATBIB.
%%% AUTHORS ARE ENCOURAGED TO USE EITHER THE "THEBIBLIOGRAPY" ENVIRONMENT
%%% BY UNCOMMENTING (DELETING THE "%" SYMBOL) THE COMMANDS BELOW, OR BY
%%% USING THE BIBTEX ENVIRONMENT. TO FIND OUT WHICH IS APPLICABLE TO YOUR
%%% CONTRIBUTION, CONSULT THE VOLUME EDITORS FOR YOUR PROCEEDINGS.
%%%
\acknowledgements

This work has been financially supported by the Special Account
for Research Grants No 70/4/9709 of the National \& Kapodistrian
University of Athens, Hellas. In the present work, the minima
database: http://var.astro.cz/ocgate/ has been used. We thank P.
Zasche for providing the Matlab code for computing the LITE
elements.

\end{document}